\documentclass[useAMS,usenatbib,usegraphicx]{mn2e}

\title[The M31 Globular Cluster System] 
  {The M31 globular cluster system: \textit{ugriz} and K-band photometry and structural parameters}
\author[M. Peacock et al.]
{Mark B. Peacock$^{1}$\thanks{E-mail:mbp@soton.ac.uk (MBP)},
Thomas J. Maccarone$^{1}$, Christian Knigge$^{1}$, 
\newauthor
Arunav Kundu$^{2}$, Christopher Z. Waters$^{3}$, Stephen E. Zepf$^{2}$, 
David R. Zurek$^{4}$\\
$^{1}$School of Physics and Astronomy, University of Southampton, Southampton, SO17 1BJ, UK\\
$^{2}$Department of Physics and Astronomy, Michigan State University, East Lansing, MI 48824, USA\\ 
$^{3}$Institute for Astronomy, University of Hawaii, 2680 Woodlawn Drive, Honolulu, HI 96822\\ 
$^{4}$Department of Astrophysics, American Museum of Natural History, New York, NY 10024, USA} 

\begin{document}

\date{Released 2009 Xxxxx XX}

\pagerange{\pageref{firstpage}--\pageref{lastpage}} \pubyear{2009}

\maketitle

\label{firstpage}

\begin{abstract}

We present an updated catalogue of M31 globular clusters (GCs) based on images from the Wide Field CAMera (WFCAM) on the UK Infrared Telescope and from the Sloan Digital Sky Survey (SDSS). Our catalogue includes new, self-consistent \textit{ugriz} and K-band photometry of these clusters. We discuss the difficulty of obtaining accurate photometry of clusters projected against M31 due to small scale background structure in the galaxy. We consider the effect of this on the accuracy of our photometry and provide realistic photometric error estimates. We investigate possible contamination in the current M31 GC catalogues using the excellent spatial resolution of these WFCAM images combined with the SDSS multicolour photometry. We identify a large population of clusters with very blue colours. Most of these have recently been proposed by other work as young clusters. We distinguish between these, and old clusters, in the final classifications. Our final catalogue includes 416 old clusters, 156 young clusters and 373 candidate clusters. We also investigate the structure of M31's old GCs using previously published King model fits to these WFCAM images. We demonstrate that the structure and colours of M31's old GC system are similar to those of the Milky Way. One GC (B383) is found to be significantly brighter in previous observations than observed here. We investigate all of the previous photometry of this GC and suggest that this variability appears to be genuine and short lived. We propose that the large increase in its luminosity my have been due to a classical nova in the GC at the time of the previous observations in 1989. 

\end{abstract}


\section{Introduction}

Globular clusters (GCs) are among the oldest known stellar systems. They typically have ages similar to those of their host galaxies, making them ideal probes into galaxy formation and evolution. The properties of GCs vary significantly. However, individual clusters contain populations of stars with similar ages and metallicities. This makes them unique locations for studying stellar evolution. 

The Milky Way's GCs still represent the best studied GC system. While the study of these clusters has led to many advances, the Milky Way contains relatively few GCs \citep[$\sim$150 GCs:][]{Harris96}, many of which have high foreground extinction, making them hard to study. By determining the properties of extragalactic GCs, we are able to study a more diverse population and ensure our current conclusions are not biased by the Milky Way's clusters being atypical. 

For extragalactic GCs, it is very difficult to resolve individual stars in the clusters. However, it is possible to estimate many properties of a GC from its integrated light. For example: the masses of GCs can be estimated by assuming a mass to light ratio; combined optical and near infrared colours of GCs can be used to (at least partially) break the age and metallicity degeneracy and estimate these parameters \citep[e.g.][]{Puzia02,Jiang03,Hempel07}; and their structural parameters can be estimated by fitting their density profiles \citep[e.g.][]{Barmby07,Jordan07,McLaughlin,Peacock09}. The colours of GCs and GC candidates are also very useful in selecting genuine GCs from stellar asterisms and background galaxies. Good multi-wavelength photometry of GCs is therefore highly desireable.

\subsection{The M31 GC system}

The proximity of M31, and its relatively large GC population compared with the Milky Way \citep[$\sim$400$:$][]{Barmby00}, makes it the ideal location to study extragalactic globular clusters. Its clusters have been the focus of many studies dating back to the early work of \citet{Hubble32} and \citet{Vetesnik62}. However, attempts to study its clusters have faced several challenges. Photometry of these clusters is complicated by many of them being projected against the bright and non-uniform structure of M31 itself. The galaxy's proximity also results in the GC system extending over a wide region of the sky, with clusters recently found beyond 4$^{\circ}$ from the centre of the galaxy \citep{Huxor08}. This means that surveys with large fields of view are required in order to study the GC system. It is also difficult to confirm GCs in M31 based on spectroscopy alone. This is because the velocity distribution of its GC system overlaps that of Milky Way halo stars. 

Over the past decades there have been several large catalogues of M31's GCs including those of: \citet{Battistini87,Barmby00,Galleti04,Kim07}. In addition to these catalogues many new clusters and candidates have been proposed \citep[e.g.][]{Battistini93,Mochejska98,Barmby02,Galleti06,Galleti07,Huxor08,Caldwell09}. These studies have made considerable progress in removing contamination from the cluster catalogues due to either background galaxies \citep[e.g.][]{Racine91,Barmby00,Perrett02,Galleti04,Kim07,Caldwell09} or stars and asterisms both in the Milky Way and M31 itself \citep[e.g.][]{Barmby00,Cohen05,Huxor08,Caldwell09}. However, despite this work, it is likely that there remains significant contamination in the current catalogues of M31 clusters, especially at the faint end of the GC luminosity function. These studies have also resulted in a large number of unconfirmed candidate clusters \citep[currently over 1000:][]{Galleti04} whose true nature remains uncertain. 

It has been known for many years that some of the proposed GCs in M31 have very blue colours. Recent work has identified that a large number of the clusters in the current catalogues are young clusters \citep[e.g.][]{Beasley04,Fusi_Pecci05,Rey07}. A comprehensive catalogue of young clusters in M31 has recently been published from the spectroscopic survey of \citet{Caldwell09}. Compared with the young open clusters in the Milky Way, these clusters have relatively high masses ($<10^{5}M_{\sun}$), akin to the young clusters observed in the Large Magellanic Cloud. A recent \textit{Hubble Space Telescope (HST)} study of 23 of these young clusters suggested that on average they are larger and more concentrated than typical old clusters \citep{Barmby09}. Most of the clusters studied by \citet{Barmby09} are found to have dissolution timescales of less than a few Gyr and are therefore not expected to evolve into typical old globular clusters. Whether these clusters are massive open clusters, young globular clusters or a mix of both, it is clear that they represent a different population to the classical old GCs also observed in M31 (which are the focus of this study). We therefore distinguish between the two populations in our classifications and conclusions. 

\begin{figure*}
 \includegraphics[width=170mm,angle=0]{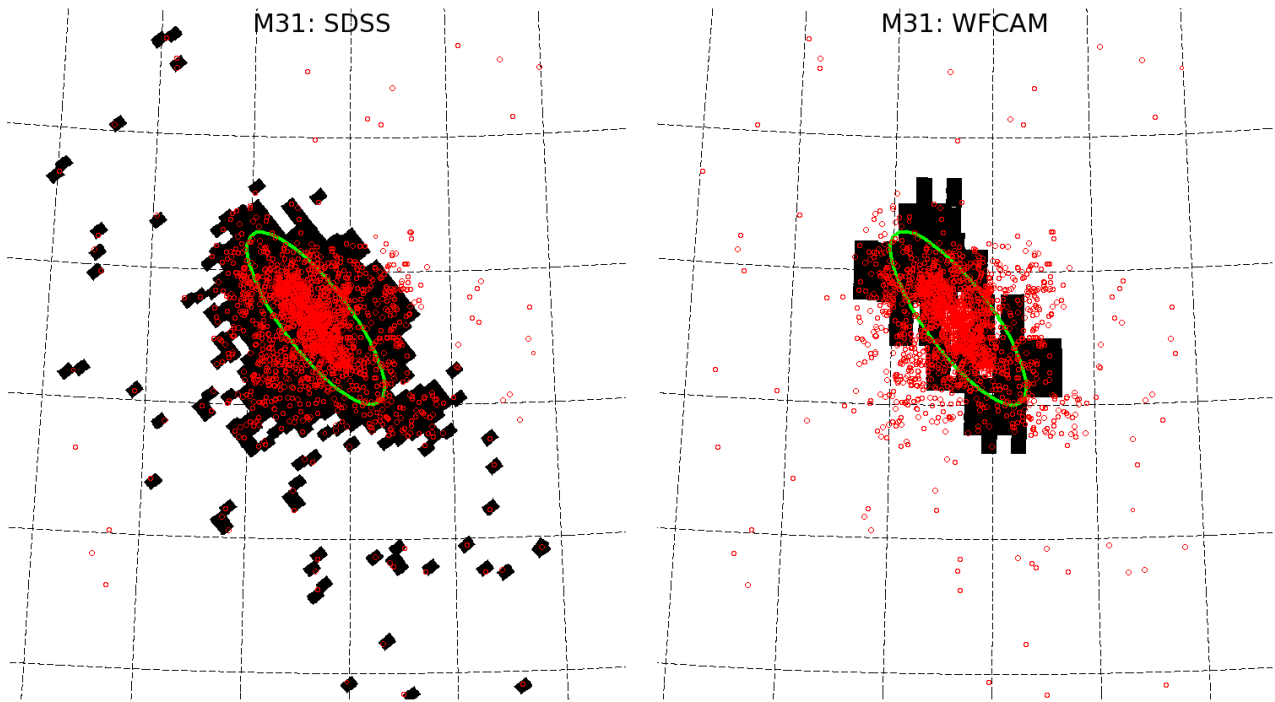}
 \caption{Coverage of the SDSS and WFCAM images used. For reference all objects listed in the RBC are shown in red. Only images covering the locations of these objects were extracted from the SDSS archive. The green ellipse indicates the D$_{25}$ ellipse of M31. The grid represents $2^{\circ}\times2^{\circ}$ squares on the sky ($13.6\times13.6$~kpc at the distance of M31) and highlights the spatial extent of the GC system. }
 \label{fig:images}
\end{figure*}

While these previous studies have provided a wealth of information on the M31 GC system they have also resulted in a rather heterogeneous sample. For example, the excellent and commonly used Revised Bologna Catalogue (hereafter RBC) of M31 GCs by \citet{Galleti04} includes photometry from many different authors using different telescopes and in some cases different (homogenised) filters. This has been previously noted by \citet{Caldwell09} (hereafter C09) who recently published new V-band photometry for many RBC sources which were located in the Local Group Galaxy Survey images of M31 \citep[LGGS:][]{Massey06}. While this work provides excellent deep V-band photometry, the survey does not cover the outer clusters and candidates and does not provide colour information. The most complete set of optical colours, derived in a consistent manner, are still that of the Barmby catalogue \citep{Barmby00}. This work presented self-consistent UBVRI colours for many of their clusters. However, it is incomplete in some of these bands and only provides new photometry 285 clusters. For these reasons we chose to produce new, self-consistent, optical photometry for the proposed GCs and GC candidates in the RBC using images from the Sloan Digital Sky Survey (SDSS). The excellent calibration and large field of view of this survey is ideal for studying such an extended system. Details of this photometry are presented in section 2.2. 

The study of M31's GCs in the near infrared (NIR) is very useful both for confirming genuine GCs and for estimating their ages and metallicities \citep[e.g.][]{Barmby00,Galleti04,Fan06}. The first major survey of M31's GCs in the NIR was by \citet{Barmby00} who used pointed observations of individual clusters to obtain K-band photometry of 228 of their clusters. More recently \citet{Galleti04} obtained NIR photometry in the J,H and K-bands of 279 of their confirmed GCs from the 2 Micron All Sky Survey (2MASS). The spatial coverage of 2MASS makes it ideal for such a project. However, the survey is relatively shallow and has relatively poor spatial resolution. We have obtained new deep K-band photometry using the Wide Field Camera on the UK Infrared Telescope to determine the K-band magnitude of M31's GCs across the entire GC luminosity function. Some results of this survey are already published in \citet{Peacock09}. In addition to providing the first K-band photometry for 126 GCs marked as confirmed in the RBC, the excellent spatial resolution of these images is very useful for removing stellar sources from genuine clusters, and for investigating the density profiles of the clusters. Details of this new K-band photometry are presented in section 2.3, while the classifications of the proposed clusters and candidates are considered in section 3. 

The proximity of M31 makes it the ideal location for studying the structure of extragalactic GCs. Determination of the size and density of GCs is very useful in investigating: stellar evolution; galaxy formation and evolution; constraining N-body simulations; and investigating exotic objects in GCs like X-ray binaries, blue stragglers and horizontal branch/extreme horizontal branch stars. The structural parameters for some of M31's GCs have been measured using \textit{Hubble Space Telescope (HST)}: Faint Object Camera (FOC) images of 13 clusters \citep{Fusi_Pecci94}; Wide Field Planetary Camera (WFPC2) images of 50 clusters \citep{Barmby02}; and Advanced Camera for Surveys (ACS) and Space Telescope Imaging Spectrograph (STIS) images of 15 and 19 clusters respectively \citep{Barmby07}. In \citet{Peacock09} we presented the results of fitting PSF convolved King models \citep{King66} to the ground based WFCAM images used here in order to estimate the structural parameters for 239 clusters. In section 4 we discuss the structure of M31's GCs.

\section{Photometry of GCs and candidates}

\subsection{Identification of GCs}

In the following analysis we consider all the GCs and GC candidates listed the RBC (their class 1$/$8 and 2 objects respectively). Based on the original catalogue of \citet{Battistini87}, this catalogue has been regularly updated to include the results from most new studies. This version of the catalogue (v3.5) includes the newly discovered GCs in the outer regions on M31 \citep{Mackey06,Huxor08} and the new GCs and candidates from \citet{Kim07} (hereafter K07: their class A and B$/$C objects respectively). We also consider the catalogue of C09 which includes some additional clusters and gives updated locations and classifications for many of the objects in the RBC based on images from the LGGS or Digital Sky Survey and/or Hectospec spectroscopy. This combined catalogue is used to identify the known GCs and candidates in the following analysis. 

\subsection{Optical photometry} 

\subsubsection{\textit{ugriz} data}

To obtain self-consistent optical photometry of M31's clusters and candidates we extracted images of M31 from the SDSS archive. Since M31 is at a relatively low Galactic latitude of -21$^{\circ}$, it is not included in the standard survey field. However drift scan images of M31 were obtained by the SDSS 2.5m telescope \citep{Adelman-McCarthy} in 2002 as part of a special run during a period when the survey's primary field was not available \citep{Zucker04}. The runs used (3366, 3367, 6426 and 7210) provide images in the five SDSS bandpasses \citep[\textit{ugriz}:][]{Fukugita}. Each of the observations takes images in these bands simultaneously meaning that they are taken under the same atmospheric conditions. The seeing for different observations varied significantly between 1.1-2.1$\arcsec$ in \textit{g} (meaning that faint GCs could appear as point sources in some of these images). The 3$\sigma$ detection limits of these images were verified to be similar to the standard survey ($u<22.0, g<22.2, r<22.2, i<21.3, z<20.5$). These data were found to cover and detect 92$\%$ (\textit{gri}), 90$\%$ (\textit{z}) and 73$\%$ (\textit{u}) of the 1558 clusters and candidates in the current RBC. Two of these GCs were saturated in the \textit{r} and \textit{i} bands and one was saturated in the \textit{g}-band. We do not provide new photometry for these clusters but good photometry is already available for these very bright clusters from previous studies. 

We extracted all images covering the locations of confirmed and candidate clusters from the SDSS Supplemental Archive. Figure \ref{fig:images} shows the coverage of these data and demonstrates that most known clusters and candidates (red circles) are covered. These images have been processed through the standard SDSS pipeline \citep{Stoughton} which both reduces the raw images and produces a catalogue of sources in each image. Since the SDSS extraction and photometry routines are not designed to work in crowded fields (like M31), the default catalogues produced by the pipeline can not be used for photometry of the clusters. Instead we performed photometry on the images as described in the next section. 

The photometric zero points for these images were calculated using the calibration coefficients produced by the pipeline. These calibrations place the magnitudes on the AB photometric system [\citet{Oke83}; the \textit{u}-band zeropoint has previously been found to be slightly offset from the AB system by $u_{\rm{AB}} = u_{\rm{SDSS}} - 0.04$~mag \citep{Bohlin01}, this correction is \textit{not} applied to our photometry]. This calibration is known to give magnitudes accurate to $\sim$0.01~mag.

\subsubsection{Identification and locations of GCs}

\begin{figure}
 \includegraphics[height=84mm,angle=270]{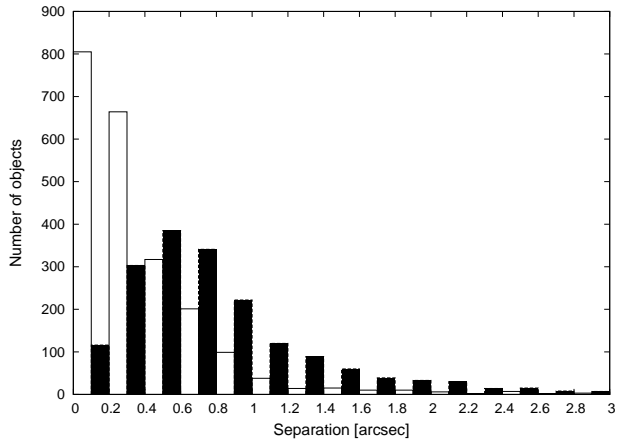}
 \caption{Difference between the location of objects in our images and their location in the RBC (solid) and C09 (open). In both cases, data are grouped into 0.2$\arcsec$ bins. }
 \label{fig:pos_error}
\end{figure}

Catalogues of all sources in each of the \textit{ugriz} images were produced using the program SExtractor \citep{Bertin96}. This detected and located every source in each filter, performed initial aperture photometry, and gave an estimate of the stellarity of each source based on the PSF of its host image. Sources were identified using a minimum detection area (DETECT$\_$MINAREA) of 3~pixels and a detection (DETECT$\_$THRESH) and analysis threshold (ANALYSIS$\_$THRESH) of 1.5$\sigma$. This threshold was chosen to ensure that the majority of the clusters profiles were included in the analysis and to help to separate genuinely extended objects from merged sources. 

This source catalogue was matched to our combined catalogue of known M31 GCs and candidates (described in section 2.1) based on astrometry. We identified all objects within 3$\arcsec$ of the locations quoted in the RBC and (separately) to their locations in C09. Some genuine clusters in the SDSS images may not appear extended due to the poor angular resolution of some of the images. Also, we wish to provide photometry for potentially misclassified stars in addition to the extended clusters. The M31 GC catalogue was therefore matched to sources with stellar profiles in addition to those with extended profiles. In the few cases where multiple sources were located within 3$\arcsec$ of the quoted location, priority was given first to sources flagged as extended and then to the closest source to the quoted location. 

Figure \ref{fig:pos_error} shows the difference between our positions and the positions quoted in RBC (solid) and C09 (open). We find excellent agreement between our locations and those of C09 but find that the difference in the positions of many sources in the RBC are greater than 1$\arcsec$. The errors in the positions of some sources in the RBC were noted and discussed by C09. We note the strong agreement between our locations and those of C09 and use their locations to identify GCs and candidates.

\subsubsection{Photometry}

Photometry of all clusters and candidates was obtained using SExtractor's simple aperture photometry. We also considered using the IRAF:APPHOT routines to perform the aperture photometry but SExtractor was found to deal better with contamination from neighbouring sources. This is a significant problem when using aperture photometry to obtain magnitudes of extended sources in a crowded region like M31. To minimise the effects of neighbouring sources within the GC aperture, SExtractor masks all other sources detected in the aperture and replaces them with pixels from symmetrically opposite the source. 

For background estimation we considered the use of both local and global solutions. To produce a global estimate of the background SExtractor produces a smoothed background map for each image. We chose to create this with a BACK$\_$FILTERSIZE of 3 and a BACK$\_$SIZE of 64~pixels. By examination of the background maps produced by SExtractor, this method was found to give a good estimation of genuine background variation (due mainly to structure in M31 itself) without subtracting flux from the sources of interest. This was compared with the photometry produced using local backgrounds (calculated around the isophotal limits of the sources). In most cases good agreement was found between the two methods. However the local background estimates were found to deal better with the most strongly varying background regions (near the center of the galaxy and its spiral arms). For this reason local background estimation was used for the final photometry. 

To determine the total luminosity of each cluster, we produced curves of growth from \textit{g}-band photometry obtained through apertures with a radii in the range $2.8-10.6\arcsec$ with 0.6$\arcsec$ increments. These were used to determine the aperture size required to enclose the total cluster light. The best aperture was determined independently for each object. This method ensures that we measure the total cluster luminosity correctly for the largest clusters. While the use of smaller apertures for smaller clusters maximises the signal to noise and minimises the contamination from nearby sources. The aperture size used to determine the total magnitude of each cluster is quoted in table 1. The average aperture radius used was $\sim$5.8$\arcsec$, with 87$\%$ of the apertures $\leq$8.2$\arcsec$. The \textit{ugriz} colours of the clusters were measured through 4$\arcsec$ apertures. We also measured the colours using the aperture determined for the total \textit{g}-band magnitude. This confirmed that there were no significant aperture effects due to the use of smaller apertures. For the final colours we chose to use the smaller aperture size in order to maximise the signal to noise and minimise the contamination from nearby sources. 

The statistical errors across the GC luminosity function are in most cases less than 0.05~mag in \textit{gri}. In general the \textit{u} and \textit{z} bands have slightly larger errors as they have slightly lower signal to noise. However, there are additional systematic errors which need to be considered. 

Firstly, the errors on the zero point calculation are estimated at 0.01~mag. This error dominates over the statistical errors for many of the bright clusters. The other significant source of error for some of the clusters is due to contamination from nearby sources and the error on the background estimation. In the bluer wavelengths there is significant small scale structure in M31. For clusters projected against the densest regions of the galaxy, this makes background subtraction difficult as it can vary on scales smaller than the cluster of interest. The issue of background estimation is found to be particularly significant for the \textit{g}-band photometry. In this filter there is significant small scale structure across M31 (this structure is less significant in the shallower \textit{u}-band images where the statistical errors are larger). In order to estimate the error on our background estimation we repeated our photometry with apertures 1.8$\arcsec$ larger than the aperture used for the total magnitudes. With perfect background estimation, the determined flux should be the same through both apertures (within the photometric errors). The difference in luminosity of the cluster through each aperture can therefore be used to give an estimation of the error on the background estimation. In most cases this estimated error is quite small, with a median value $\sim$0.015~mag, but for a few clusters it can reach $\sim$0.1~mag. This additional error is combined with the calibration and statistical error and included in table 1 as $\sigma_{g,tot}$. 

The error on the \textit{ugriz} colours should be less affected by these effects. This is because they are often taken through smaller apertures, also we expect that possible errors on the background level in each filter should, at least partially, cancel. For this reason the errors on the colours quoted in table 1 are only the statistical errors.

\subsubsection{Comparison with previous photometry}

\begin{figure}
 \includegraphics[height=84mm,angle=270]{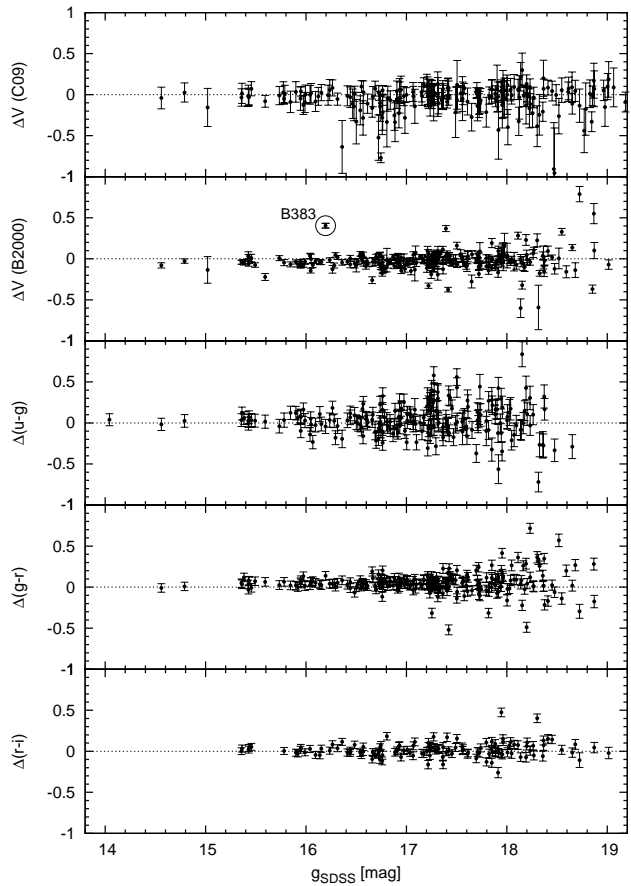}
 \caption{\textit{Top:} comparison between our total cluster magnitudes and those from C09. \textit{Bottom:} comparison between our total cluster magnitudes and colours and those from the Barmby catalogue (B2000). In all cases the y axis is our photometry minus that obtained previously. The highlighted point indicates the cluster B383.}
 \label{fig:compare_ugriz}
\end{figure}

\begin{table*}
\setcounter{table}{1}
 \centering
 \begin{minipage}{160mm}
  \caption{Photometry of B383 \label{tab:B383}}
\end{minipage}
 \begin{minipage}{160mm}
  \begin{tabular}{@{}llcrrrrl@{}}
  \hline
  \hline
   Source of photometry & Observation date & Detector  & $\Delta$U & $\Delta$B & $\Delta$V & $\Delta$R &\\
  \hline
   SDSS (This study)    & 2002 October 06    & CCD           &[ 17.15 & 16.70 & 15.72 & 15.06 &]$^{\star}$\\
   \citet{Sharov92}     & 1990 October 15-18 & Photoelectric & -0.13 & -0.06 & -0.06 & - &\\
   \citet{Reed92}       & 1989 August 23-30  & CCD           & -     &  0.54 &  0.39 & 0.18 &\\
   \citet{Sharov83}     & 1980 October 8-13  & Photoelectric & -0.05 & -0.03 & -0.06 & - &\\
   \citet{Battistini87} & 1977-1981          & Plate         &  0.16 &  0.07 &  0.16 & - &\\
  \hline
  \end{tabular}\\
The magnitude difference between the SDSS photometry presented here and that found by previous work. \\
$^{\star}$ The total magnitude of the cluster in the SDSS images (transformed to Johnson filter system) 
\end{minipage}
\end{table*}

To date the best set of optical colours of M31 GCs derived in a consistent manner is that of the Barmby catalogue. This includes colours for 285 clusters in the Johnson UBVRI bands obtained through 8$\arcsec$ apertures. The catalogue also contains photometry for an additional 160 clusters collated from other studies. This collated photometry is taken mainly from the work of: \citet{Reed92,Reed94,Battistini93,Mochejska98,Sharov83,Sharov87,Sharov92,Sharov95}. For full details of the sources and reliability of this additional photometry we refer the reader to the description in \citet{Barmby00} and the references therein. 

There is little previous optical photometry in the \textit{ugriz} bands with which to compare our results. However it is possible to compare our colours with those of the Barmby catalogue by transforming between the UBVRI and \textit{ugriz} bands. This was done using the following the transformations from \citet{Jester}
\begin{equation}
 V   = g - 0.59\times(g-r) - 0.01    \pm0.01
 \label{eq:trans_V}
\end{equation}

\begin{equation}
 u-g = 1.28\times(\rm{U-B})   + 1.13  \pm0.06
 \label{eq:trans_ug}
\end{equation}

\begin{equation}
 g-r = 1.02\times(\rm{B-V})   - 0.22  \pm0.04
 \label{eq:trans_gr}
\end{equation}

\begin{equation}
 r-i = 0.91\times(\rm{Rc-Ic}) - 0.20  \pm0.03
 \label{eq:trans_ri}
\end{equation}
These transformations are based on all stars studied by \citet{Jester}. Applying these to the colours of globular clusters may introduce a slightly larger error than the quoted rms residuals as globular clusters are stellar populations rather than single stars. However, they can be used to check for consistency with this previous work.

Figure \ref{fig:compare_ugriz} compares our colours for confirmed clusters with the transformed colours from the Barmby catalogue. The errors quoted include the residual from the transformations and the errors on our photometry only. The scatter is therefore expected to be larger than 1$\sigma$ due to errors on the previous photometry. It can be seen that reasonable agreement is found between the \textit{u-g} and \textit{r-i} colours of the clusters. For the \textit{g-r} colours a slight offset of 0.035~mags is found. However, this is within the rms scatter of the transformations. We believe this offset may be due to the errors in the transformations (due to the difference in the spectrum of a typical globular cluster compared with a single star), rather than a genuine offset between the colours. We therefore believe that for most clusters our colours are consistent with the previous UBVRI colours in the Barmby catalogue. 

The top panels of figure \ref{fig:compare_ugriz} compares the total magnitudes of the clusters obtained here with V-band photometry from the Barmby catalogue and the more recent photometry of C09. To compare the total magnitudes of the clusters, our \textit{g}-band photometry was transformed to the V-band using equation \ref{eq:trans_V}. For most clusters good agreement is found between our magnitudes and those in the Barmby catalogue (again errors on the Barmby catalogue photometry are not included). However, it can be seen that there are some significant outliers. Some of the brightest clusters are brighter in our photometry than found previously. This is likely due to our use of larger apertures for larger clusters. Many of these bright clusters are found to extend beyond the 8$\arcsec$ apertures used to obtain the Barmby catalogue photometry. For the fainter clusters we identify a group of 7 clusters which are fainter than expected. These clusters all have nearby neighbours, the effects of which we attempt to remove from our photometry but believe are included in the previous photometry. We therefore believe our values for these clusters to represent the actual cluster magnitudes better. Another group of faint clusters are found which are brighter than expected. The majority of these clusters are in dense regions near the galaxy center or spiral arms and we believe the differences are due to errors in the background estimation. It is very difficult to estimate the background accurately for regions with variations on the scales of the clusters themselves. It is unclear which photometry is more accurate for these few clusters, although our use of smaller apertures for smaller clusters should minimise this effect. In the Barmby catalogue they subtract light from the bulge of M31 before performing photometry using a ring median filter. We repeated our photometry using a similar method but did not find significant differences in our photometry. Background estimation for clusters in these dense regions is an inherent problem in finding their absolute magnitudes. It should be noted that, while we attempt to account for this in the quoted errors on our photometry, the errors for some clusters in these dense regions may be larger than quoted. 

It can also be seen that excellent agreement is found between between our photometry and that of C09. The errors on this comparison are larger due to the inclusion of the errors quoted by C09. The group of clusters which were fainter in our photometry than the Barmby catalogue are found to agree well with this photometry. This is likely due to C09 also subtracting the effects of nearby sources from their photometry. They also use a similar method of increasing their aperture size for larger clusters, and our photometry for brighter clusters agrees with theirs. We again identify a few clusters in dense background regions whose magnitudes are slightly fainter than expected. However, not all of these are the same outliers as those found in the Barmby catalogue. This highlights the difficulty of accurately obtaining integrated magnitudes for clusters in these regions. 

\subsubsection{Variability in B383: a classical nova?}

Figure \ref{fig:compare_ugriz} identifies one relatively bright cluster (B383) which is significantly fainter ($\Delta\rm{V}=0.39$) in our photometry than found in previous photometry. This cluster was not observed by \citet{Barmby00} and its BVR band photometry in both the Barmby catalogue and the RBC are from the work of \citet{Reed92}. This cluster has high signal to noise, a relatively clean background, and its magnitude was obtained through a similar sized aperture to that used previously (7.8$\arcsec$). The cluster is present in two different SDSS observations and the magnitudes obtained from each agree very well. Table \ref{tab:B383} compares our photometry with other previous observations of B383. It can be seen that there is good agreement between our photometry and the previous photometry of \citet{Sharov83}, \citet{Sharov92} and \citet{Battistini87}. We therefore believe that our photometry of this cluster is reliable. 

We note that, for other clusters our photometry agrees well with that of \citet{Reed92} and that B383 is brighter in all of their observations (B, V and R bands). We therefore believe that this discrepancy is unlikely to be due to an error in their photometry. This raises the possibility that the cluster luminosity may have genuinely varied between our observations. The increase in luminosity of B383 could have been produced by a transient in the cluster. To explain the observed variability, this transient would have to have bluer colours than the cluster and a brightness of M$_{\rm{V}}\sim-7.9$. 

A potential candidate for this increase would be a classical nova in the cluster at the time of the \citet{Reed92} observations. Novae have typical luminosities of $-6 < M_{\rm{V}} < -9$ and could explain this blue excess in the \citet{Reed92} observations. A classical novae of this brightness would be expected to have a very short outburst duration \citep[e.g.][]{Warner95} and would therefore be expected to have faded by the time of our observations and even those of \citet{Sharov92} $\sim$10 months later. 

Globular clusters are expected to host classical novae. There is evidence for Classical novae in the Galactic GCs M80 \citep{Pogson1860,Wehlau90} and (possibly) M14 \citep{Hogg64,Margon91}. Classical novae have also been detected in a GC in M87 \citep{Shara04} and two of M31's other GCs [B111: \citet{Quimby07,Shafter07} and B194: \citet{Henze09}]. Confirmation of a classical nova in B383 is very difficult as any remaining signatures of the event will be very faint. However, it offers a plausible explanation for such a large brightness variation.

\subsection{NIR Photometry}

\subsubsection{K-band data}

To obtain K-band photometry of M31's GCs, images across M31 were obtained using the Wide Field CAMera (WFCAM) on the UK Infrared Telescope (UKIRT) under the service program USERV1652. The large field of view of the WFCAM makes it ideal for such a project. The coverage of these data is shown in figure \ref{fig:images}. They do not currently cover the whole GC system, missing both the most central and most distant clusters. The details of these observations were originally presented in \citet{Peacock09} but are summarised again below. 

The data were taken on the nights of 2005 November 30 and 2007 August 06 with K-band seeing of 0.85-1.00$\arcsec$ and 0.6-0.8$\arcsec$ respectively. To ensure the images were well sampled, each observation was taken with 2$\times$2 microstepping to give an effective pixel size of 0.2$\arcsec$. Five observations were taken of each field giving a total exposure time of 225s and a 3$\sigma$ detection limit of $\sim$19~mag. 

The images were reduced using the standard WFCAM pipeline \citep[see e.g.][]{Dye}. The pipeline processing reduced and stacked the raw images and interlaced \citep{Fruchter02} the microstepped images together. The pipeline also applies an accurate astrometric solution to the images based on matching sources to the 2MASS catalogue. This method has been shown to give positions accurate to 80mas \citep{Dye}. We determined the photometric zero point for each observation by calibrating against the 2MASS catalogue. This was done by comparing instrumental magnitudes of bright, unsaturated, stars in each field with the 2MASS Point Source Catalogue. This places the K-band photometry on the standard 2MASS (Vega-based) photometric system. This method has previously been shown to give zero points for K-band WFCAM images to better than 0.02~mag \citep{Hodgkin09}. 

\subsubsection{Photometry} 

Photometry was obtained for all GCs and candidates with WFCAM images using SExtractor. SExtractor was run in the same way used to obtain the optical magnitudes (described in section 2.2). The aperture required to determine the total K-band luminosity was again selected for each cluster from curves of growth. The aperture size was selected independently of the aperture used to determine the total \textit{g}-band magnitude of the cluster. In many cases a smaller aperture was required to enclose all the K-band light of the cluster, with an average aperture size of $\sim$4.6$\arcsec$ selected and 84$\%$ of the apertures $\leq$~6$\arcsec$. The use of smaller apertures for the K-band images is expected because of the smaller PSF of these images. 

As with the \textit{ugriz} photometry, it was found that the error in the K-band photometry was often dominated by non statistical errors. The zero point calibration error of the WFCAM images is estimated to be 0.02~mag. This is larger than the statistical errors for most of the clusters. The K-band luminosity of the clusters also suffers from errors on the background estimation and contamination from neighbouring sources. We estimate the effect of this on the accuracy of our photometry using the same method used for the total \textit{g}-band magnitude (by retaking photometry through a larger aperture). The median estimated error, due to contamination, was found to be 0.025~mag. The estimated error due to contamination and background variation was combined with the statistical and calibration errors and quoted as the final error $\sigma_{K,tot}$ in table 1. 

\subsubsection{Comparison with previous photometry}

\begin{figure}
 \includegraphics[height=84mm,angle=270]{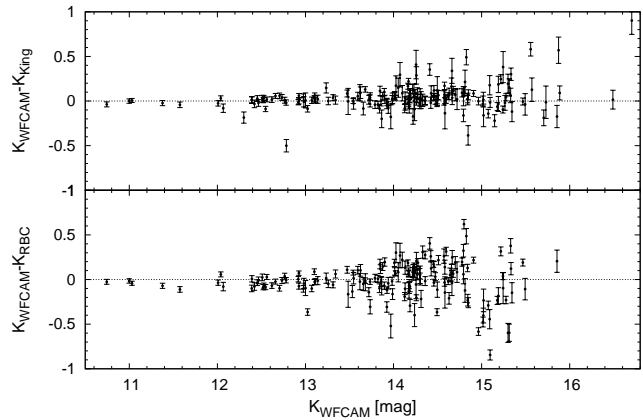}
 \caption{Comparison with previous K-band photometry from the RBC (\textit{bottom}) and profile fits to these WFCAM images from \citet{Peacock09} (\textit{top}). The errors quoted are from our photometry only, as errors are not available from the previous work.}
 \label{fig:compare_k}
\end{figure}

The most complete NIR data currently available for M31's GCs is that from the RBC. This includes K-band magnitudes of 279 confirmed GCs in the RBC obtained from the 2MASS archive \citep{Galleti04}. This 2MASS photometry is from either the 2MASS Point Source Catalogue or Extended Source Catalogue and measured through apertures with radii of 4 and 5$\arcsec$ respectively. The bottom panel of figure \ref{fig:compare_k} compares our K-band photometry of all confirmed GCs with the K-band photometry in the RBC obtained from 2MASS. Errors are not included for the K-band photometry in the RBC, so only the total errors on our photometry are included in these plots. 

\begin{figure}
 \includegraphics[width=84mm,angle=0]{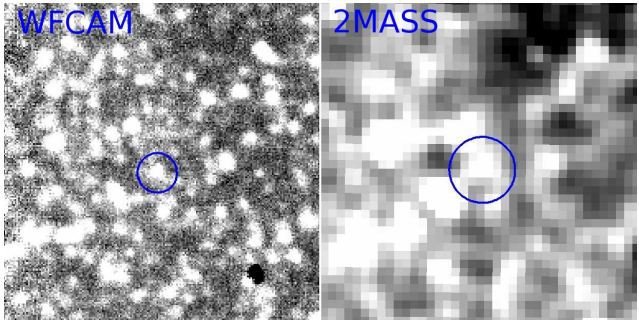}
 \caption{WFCAM and 2MASS images of B041. Both images are $45\arcsec\times45\arcsec$ and demonstrate the improved spatial resolution and signal to noise of the WFCAM images over 2MASS.}
 \label{fig:B041}
\end{figure}

Two clusters are found to have very different magnitudes and lie off this plot. One of these (B090) is very faint in the previous photometry and has a very blue J-K colour. We therefore believe the previous photometry for this object is unlikely to be accurate. The other (B041) is found to be fainter in our photometry. The WFCAM and 2MASS images for this cluster are shown in figure \ref{fig:B041}. This comparison demonstrates the superior depth and spatial resolution of the WFCAM images over 2MASS. The circles show the aperture sizes used for our photometry and the 2MASS photometry. The improved spatial resolution helps to separate clusters from nearby sources and allows the use of significantly smaller apertures. It is clear from this image that the previous photometry for these faint sources is unlikely to be as reliable as that presented here. From examination of the 2MASS catalogues we can only identify this source in the `reject' catalogue. We therefore believe that the 2MASS photometry of this cluster is unreliable. 

The errors on the 2MASS photometry are expected to be larger than those obtained here since 2MASS is significantly shallower than our data. Taking this into account, most of our photometry is found to be consistent, although there are several outliers. We believe that most of these differences are due to the improved spatial resolution of our data over 2MASS which makes it easier for us to remove contamination from nearby sources and to estimate the background level more reliably. The poorer resolution could result in both overestimation of the cluster magnitudes (if nearby sources are included in the cluster aperture) and underestimation of their magnitudes (if unresolved background sources result in an overestimation of the background level). This highlights the importance of spatial resolution, even in obtaining integrated magnitudes. The brightest clusters are also found to be $\sim0.05$ mag brighter in our photometry than in the RBC. We believe this is due to our use of larger apertures for larger clusters. This was identified and discussed by \citet{Galleti04} who attempt to apply aperture corrections to these clusters. However, we believe our use of larger apertures should be more reliable. 

An alternative method to aperture photometry is to fit the profile of the clusters and find their integrated magnitudes. This method removes aperture affects because it integrates the magnitude out to the tidal radius of the cluster. It also accounts for contamination from nearby sources since it assumes the cluster to have a smooth profile. This provides a very useful independent method of estimating the total magnitudes of the clusters. The results of fitting the profiles of clusters in these images are already presented in \citet{Peacock09}. The top panel of figure \ref{fig:compare_k} compares this integrated K-band magnitude with the aperture magnitudes found here. Again errors are not available for the profile fit magnitudes but are expected to be of a similar size to the errors obtained from aperture photometry. Three of these clusters are found to be brighter in our photometry. Examination of these clusters revealed that they all have very bright nearby neighbours. Since these cause significant background gradients across the cluster profiles, we believe that the model fits to these will be less reliable and the aperture photometry is probably more accurate. Some of the fainter clusters are also found to lie slightly outside 2$\sigma$. As discussed later, we believe this is due to the King model fits being less reliable for these faint clusters. For these faintest clusters it is likely that aperture photometry gives more accurate magnitudes. 

The scatter in these comparisons highlights the difficulty in determining the NIR magnitudes of clusters projected onto stars and surface brightness fluctuations from M31. We believe our approach gives the best estimate of their magnitudes and the most realistic errors to date. In total we present K-band photometry for 319 and 603 sources classified as confirmed and candidate clusters in the RBC respectively. This includes the first K-band photometry for 126 confirmed clusters and 429 candidate clusters.

\subsection{Summary of photometry} 

\begin{figure*}
 \vspace{-8mm}
 \hspace{-24mm}
 \includegraphics[width=184mm,angle=180]{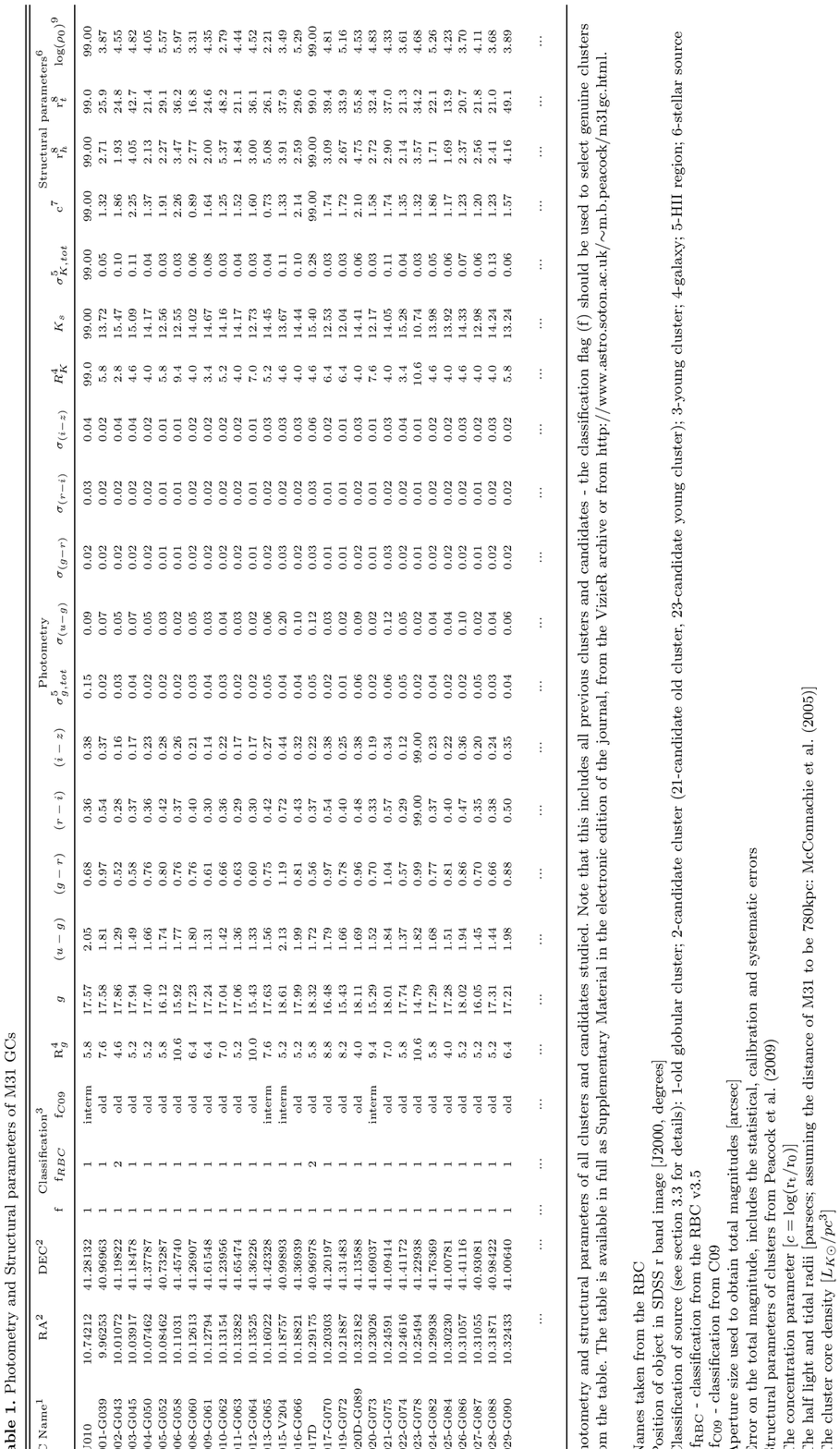}
 \label{*}
\end{figure*}

\begin{figure*}
 \includegraphics[height=178mm,angle=270]{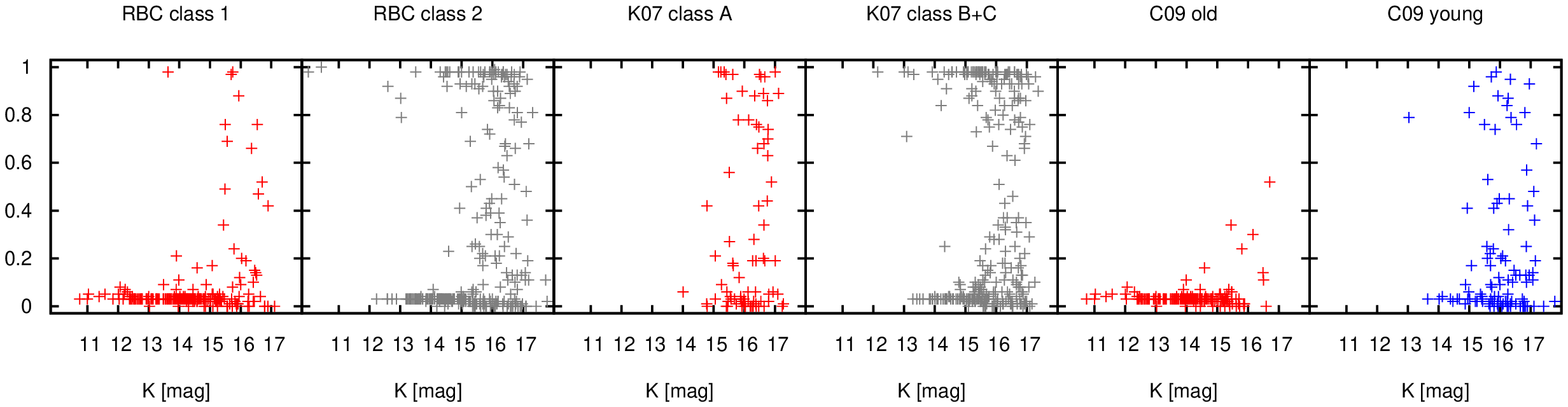}
 \caption{Stellarity of objects classed as confirmed clusters (red) and candidate clusters (grey) in the RBC (their class 1 and 2 sources respectively) and from K07 (their class A and B$/$C sources respectively). Also included are the re-classifications for many of these sources from C09 (their old and young clusters). The stellarity is based on SExtractor photometry of the WFCAM images. Extended sources have a stellarity close to 0 and point sources close to 1.}
 \label{fig:stellarility}
\end{figure*}

The \textit{ugriz} colours, total \textit{g} and total K-band luminosity of M31's GCs and candidates are presented in table 1. This table includes the statistical errors on the \textit{ugriz} colours and the errors on the total \textit{g} and K-band luminosity (which include the calibration errors and estimated error due to background variation and contamination from nearby sources). 

It should be noted that the \textit{ugriz} photometry presented here is on the standard SDSS (AB) photometric system, while the K-band photometry is on the standard 2MASS (Vega-based) photometric system. The magnitudes can be converted between the two systems using the following offsets taken from \citet{Hewett06}: 

\begin{equation}
 u_{\rm{Vega}} = u_{\rm{AB}}-0.927 
 \label{eq:AB_VEGAu}
\end{equation}
\begin{equation}
 g_{\rm{Vega}} = g_{\rm{AB}}+0.103 
 \label{eq:AB_VEGAg}
\end{equation}
\begin{equation}
 r_{\rm{Vega}} = r_{\rm{AB}}-0.146
 \label{eq:AB_VEGAr}
\end{equation}
\begin{equation}
 i_{\rm{Vega}} = i_{\rm{AB}}-0.366
 \label{eq:AB_VEGAi}
\end{equation}
\begin{equation}
 z_{\rm{Vega}} = z_{\rm{AB}}-0.533
 \label{eq:AB_VEGAz}
\end{equation}
\begin{equation}
 \rm{K_{AB} = K_{Vega}+1.900}
 \label{eq:AB_VEGAK}
\end{equation}

The names of the objects in table 1 are taken from the RBC. The positions of the sources are taken from their locations in our \textit{r}-band images and should be accurate to better than 1$\arcsec$. Some of the proposed clusters are not detected (or not located) in the SDSS images. The names, locations and classifications of these clusters (taken from the RBC or C09) are included in table 1. This table lists all previously proposed clusters and candidates in the RBC. Many of these objects are found by this (and other) studies not to be genuine clusters. The classifications of these sources are discussed in the next section. Only those objects with classification flag, f=1 should be considered confirmed old GCs.

\section{Classification of sources} 

\subsection{Stellarity}

The WFCAM images of M31 have a PSF of $0.6-0.95\arcsec$ corresponding to a spatial resolution of 2.1-3.6 pc at the distance of M31. This is a significant improvement over most of the images previously used to classify clusters. It allows us to investigate possible contamination in the previous GC catalogues from single stars and previously unresolved asterisms of stars. Figure \ref{fig:stellarility} shows the SExtractor K-band stellarity flag for confirmed and candidates clusters in the RBC (left) and, separately, the clusters and candidates from K07 (middle). Also included is the stellarity of old and young clusters from C09 (most of which are re-classifications of sources in the other two catalogues). 

The majority of sources can be identified as either having stellar profiles (with a stellarity close to 1) or extended profiles (with stellarity close to 0). It can be seen that some objects with K$>$15 have uncertain stellarity flags. The ability of SExtractor to determine the stellarity of a source is mainly dependent on the signal to noise of the source, the PSF of the image and crowding around the source. From visual examination of the sources with uncertain stellarity flags, it was found that the majority of them have nearby sources contaminating their profiles. In general we consider objects with a stellarity $<$0.4 to be extended. However, it is clear that this classification is less reliable for those objects with uncertain flags. For these objects we rely on visual examination of the cluster to estimate their nature (as described in the next section). 

It can be seen from figure \ref{fig:stellarility} that excellent agreement is found between our data and the classifications of C09 with all sources they classify as old being extended. We also find that the majority of confirmed clusters in the RBC are extended. However, there are 12 RBC \textit{class 1} objects which have either stellar or uncertain stellarity flags. We note that some of these clusters have already been reclassified by C09 as stars. It can also be seen that many of the sources classed as confirmed clusters by K07 are found to be unresolved. We note that their work was based on images with poorer spatial resolution and we reclassify many of these objects as being stellar sources. 

Some of the young clusters from C09 are found to be extended and look like normal centrally concentrated GCs. However, it can be seen from figure \ref{fig:stellarility} that many of the proposed young clusters have stellar, or uncertain, stellarity flags. This is likely because these young clusters can appear as resolved asterisms in the K-band images. This has previously been noted by \citet{Cohen05} who used K-band images taken with adaptive optics to demonstrate that 4 proposed young clusters may be asterisms. However, as discussed by C09, young clusters are generally faint in K and may be dominated by only a few bright (resolved) supergiants making them appear as resolved asterisms of stars, rather than an extended cluster. Many of these objects have subsequently been confirmed by \textit{HST} images to be genuine clusters. We therefore do not reclassify any of the proposed young clusters which appear as resolved stellar sources in our K-band images. 

Our data also allow us to classify many of the previously unclassified candidate clusters. In total we classify 368 previous candidates as likely to be stellar sources. For the above reasons, it is possible that we may potentially include some genuine young clusters in this classification. Figure \ref{fig:stellarility} demonstrates that a large group of the proposed candidates are extended in our images. These objects are therefore likely to be either genuine clusters or background galaxies. These candidates represent ideal targets for followup spectroscopy in order to confirm their nature. 

\subsection{Visual examination} 

\begin{figure*}
 \includegraphics[height=176mm,angle=270]{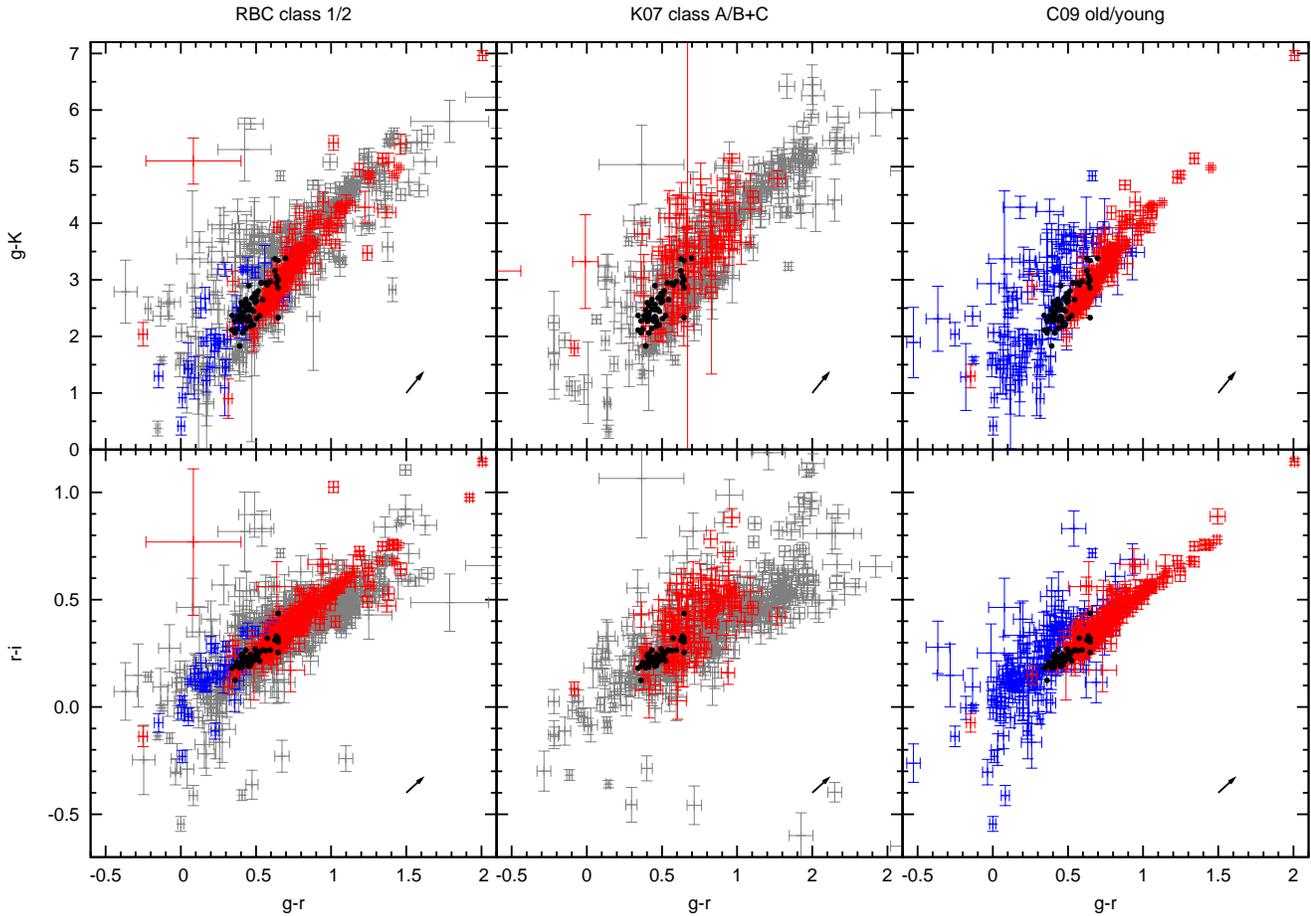}
 \caption{Colours of clusters and candidates from the RBC (\textit{left}), K07 (\textit{middle}) and C09 (\textit{right}). Included are objects classified as confirmed old clusters (red), young clusters (blue), and candidate clusters (grey) from each catalogue. The black points show the colours of Milky Way GCs. The arrow represents a reddening of $E(B-V)=0.1$.}
 \label{fig:colours}
\end{figure*}

As discussed above, the stellarity flag for some of the fainter objects is relatively uncertain. Visual examination of these objects can help in deciding whether they are extended or stellar sources. Visual examination of the clusters and candidates which are confirmed as extended also provides a method of identifying background galaxies. These were identified as either having spiral structure, or extended ellipticity. While this method is relatively subjective, it is helpful in classifying an object. Inspecting the images of the objects also provides a useful check on our otherwise automated classifications. During this process, we also ensured that our automated photometry had selected a reasonable aperture size for each cluster, in order to measure its total luminosity. 

We examined the \textit{ugriz} and K-band images of every cluster and candidate studied\footnote{Thumbnail images of these clusters are available at http://www.astro.soton.ac.uk/$\sim$m.b.peacock/m31gc.html}. We first examined the objects in our sample which have recently been classified from the spectroscopic study of C09 as being background galaxies. We then examined the previously classified confirmed clusters, followed by the proposed candidate clusters. In this way, we were able to reclassify some of the objects based on their appearance. We note that some of the newly confirmed galaxies look very similar to typical GCs. This highlights the limitations of visual examination on identifying galaxies. We do not reclassify any of the previously confirmed clusters as galaxies based on this visual examination. However, we did identify 3 candidate clusters with clear spiral structure and 30 other candidates which are likely background elliptical galaxies (this is in addition to the candidates confirmed to be galaxies from the spectroscopic study of C09). During this visual examination it was also found that some of the clusters and candidates with uncertain stellarity flags from our SExtractor photometry are likely to be multiple stellar sources, rather than extended clusters. 

\subsection{Colours}

\begin{figure*}
 \includegraphics[width=130mm,angle=270]{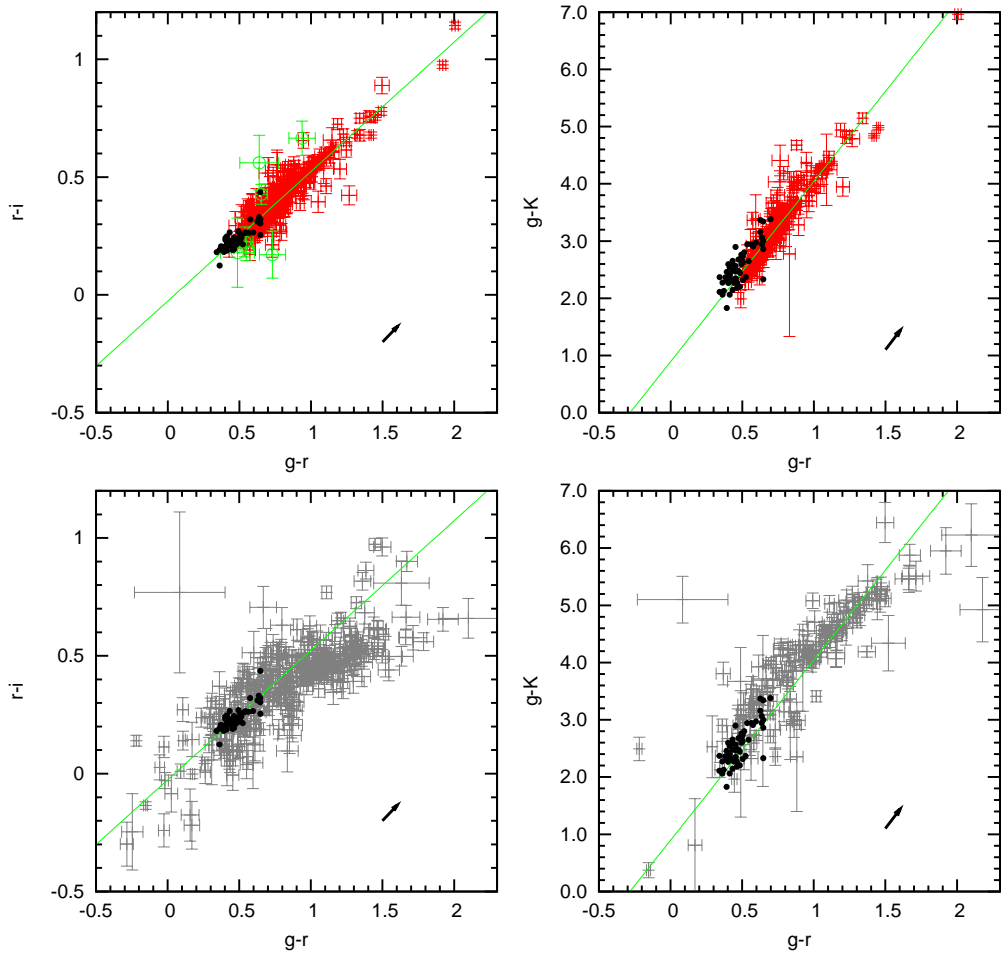}
 \caption{\textit{Top:} Colours of previously confirmed clusters which are confirmed here to be extended (red). \textit{Bottom:} Colours of proposed candidate clusters which are confirmed to be extended (grey). The lines indicate linear fits to the colours of all confirmed M31 GCs. The black points indicate the colours of the Milky Way GCs and the arrow represents an extinction of $E(B-V)=0.1$.}
 \label{fig:colours_final}
\end{figure*}

Figure \ref{fig:colours} shows the colours of objects previously classified as confirmed GCs (red) and candidate GCs (grey) in the RBC (left) and by K07 (middle). The right panel shows the colours of the objects which are confirmed by C09 to be old clusters. Shown in blue are the proposed young clusters from C09 and the confirmed clusters from the RBC which are flagged as being potentially young. For comparison, the black points indicate the colours of the Milky Way's GCs. The \textit{g}-K colour for the Milky Way GCs were taken from \citet{Cohen07} and optical colours from the Harris catalogue \citep{Harris96}. The colours of the Milky Way GCs were transformed into the \textit{ugriz} filters using the transformations of \citet{Jester} and dereddened using the values for $E(B-V)$ quoted in the Harris catalogue. Only the Milky Way clusters with $E(B-V)<0.4$ are included. This limits the Milky Way sample to mainly low metallicity clusters. It can be seen that the Milky Way's GCs define a tight region in the colour-colour plots. For this reason, the colours of the proposed GCs and candidates in M31 are very useful in classifying the objects. 

It should be noted that the colours of M31's GCs are reddened due to both Galactic extinction and extinction intrinsic to M31. The Galactic reddening in the direction of M31 is relatively uncertain, but it is estimated for the region around the disk of M31 to be $E(B-V)\sim$0.062 mag \citep{Schlegel98}. However, the extinction due to M31 itself can be much larger and varies significantly between GCs due to their locations in (and line of sight depths through) the galaxy. Previous work \citep[e.g.][]{Barmby00,Fan08} has demonstrated that the reddening for some of these clusters can be substantial. For example, the very red cluster in figure \ref{fig:colours} with \textit{g}-K=6.95 is B037 which is known to be heavily reddened [$E(B-V)=1.38$: \citet{Barmby00}]. 

Figure \ref{fig:colours} shows that our colours are in good agreement with the classifications of C09. It can be seen that most of the objects classified by C09 as old clusters define a tight region which is consistent with the (reddened) colours of the Milky Way's GCs. In most cases the confirmed clusters in the RBC also have colours consistent with the Milky Way's GCs. Many of the confirmed clusters from K07, and a few of the confirmed clusters from the RBC, have colours which are not consistent with the Milky Way's GCs or the majority of the confirmed GCs in M31. This is in agreement with our conclusions from the previous section that some of the previously confirmed clusters may be misclassified stars. The colours also suggest that many of the unclassified candidate clusters may be stars, asterisms of stars or background galaxies. 

\begin{table*}
\setcounter{table}{2}
 \centering
 \begin{minipage}{180mm}
  \caption{Classifications of sources \label{tab:compare_class}}
\end{minipage}
 \begin{minipage}{140mm}
  \begin{tabular}{@{}lrrrrrrr@{}}
  \hline
  \hline
   Classification & Number in & \multicolumn{6}{c}{Previous classification of these objects} \\
                  & this study & RBC \textit{1} & RBC \textit{2} & K07 \textit{A} & K07 \textit{B$/$C} & C09 \textit{old} & C09 \textit{young} \\
  \hline
   1: old globular cluster & \textbf{416} & 342 &  41 & 27 &   0 & 336 &   0 \\
   2: candidate cluster    & \textbf{373} &   6 & 101 &  9 & 256 &   3 &   0 \\
   3: young cluster        & \textbf{156} &  46$^{\star}$ &  78 &  2 &   0 &   2 & 151 \\
   4: background galaxy    & \textbf{189} &   5 & 170 &  4 &  10 &   0 &   0 \\
   5: HII region           & \textbf{ 17} &   0 &  14 &  3 &   0 &   0 &   0 \\
   6: stellar source       & \textbf{444} &  10 & 153 & 66 & 215 &   1 &   0 \\
  \hline
   \multicolumn{2}{r}{Total (previous catalogues):} & 409 & 557 & 111 & 481 & 342 & 151 \\
  \hline
  \end{tabular}\\
$^{\star}$ Many of these RBC \textit{class 1} clusters are flagged separately in the RBC as potentially young clusters. 
\end{minipage}
\end{table*}

\subsubsection{Young clusters}

These colours clearly identify the population of very blue clusters that have been noted by previous studies. It can be seen that our colours are in excellent agreement with the spectroscopic classifications of C09. We also find good agreement with the confirmed clusters in the RBC which are flagged as potential young clusters (flag yy$=$1,2 or 3 in the RBC). This flag is based on the work of \citet{Fusi_Pecci05}. Most of the previously identified young clusters are much bluer in \textit{g-r} than any GC in the Milky Way. Using a similar method to \citet{Fusi_Pecci05}, we define all objects with \textit{g-r}$<$0.3 to be young clusters. 

Some of the proposed young clusters have colours which are consistent with being old clusters. However, these objects are also consistent with being young clusters with reddened colours. These clusters are also found in high density regions of M31 and look similar to the other young clusters we have observed. We therefore choose to keep the previous (spectroscopic) classification for these clusters and suspect that their colours may be reddened. It can also be seen from figure \ref{fig:colours} that two clusters classified as old by C09 (B386 and PHF7-1) have very blue colours. We reclassify these two objects as young clusters.

\subsubsection{Old globular clusters} 

Figure \ref{fig:colours_final} shows the colours of all confirmed and candidate clusters following the removal of all stellar objects based on their stellarity flag or visual examination of the cluster images. We have also removed those objects identified in the previous section as being as young clusters. It can be seen that, having removed these objects, the colours of the confirmed clusters are now consistent with the colours of the old GC system of the Milky Way. The clusters extend to much redder colours, but this is consistent with the expected reddening due to extinction from M31. 

The grey points in the bottom panels of figure \ref{fig:colours_final} show the colours of the remaining candidate clusters after the removal of non-extended objects. Comparison with the confirmed old clusters shows that many of the candidates have colours consistent with being old clusters. These clusters are flagged as old candidates in table 1 and should be considered the strongest candidate clusters. We also identify candidates with very blue colours, consistent with the other young clusters identified. These are flagged as young candidate clusters in table 1. It can be seen that, despite removing objects identified as stars, the colours of many of the candidate clusters are inconsistent with being either old or young clusters. As we are uncertain of the classification of these objects, we retain their classification as candidates. However, it is likely that many of these candidates are either background galaxies or unresolved asterisms.

\subsubsection{Extended clusters in the halo of M31} 

Recent studies of the halo of M31 have identified a population of very extended clusters \citep{Huxor05,Mackey06,Huxor08}. These clusters have half light radii much greater than the majority of clusters in M31. For a description of these clusters we refer the reader to \citet{Huxor08}. Seven of these clusters are located in our SDSS images and can be identified in table 1 from their names which have the prefix HEC (`Halo Extended Cluster'). Our colours of these clusters were found to be less reliable than the other clusters studied. This is because they are resolved, due to the diffuse nature of the clusters, into multiple sources. 

The colours for these clusters were therefore re-measured through 12$\arcsec$ apertures using the IRAF:APPHOT task PHOT. A smaller aperture of 8$\arcsec$ was used for HEC11 due to a bright neighbouring star. This method gives reliable results for clusters in the halo of M31 where there is little contamination from neighbouring sources and the background is relatively smooth. None of these extended clusters are identified in the inner regions of M31, although detecting such extended and faint objects in front of the M31 would be very difficult. 

These clusters are identified in figure \ref{fig:colours_final} as open green points. It can be seen that the colours of these clusters are now consistent with the other old GCs in M31. The errors on the colours of the HEC clusters are larger than those of the other GCs. This is due to their diffuse nature and the use of large apertures, which increases the total sky background. 

\subsection{Final classification}

Our final classification is based on: the stellarity of the object; its colours; visual examination of the object in our 6 bands; velocity information and classifications from previous studies. Table 1 lists these classifications for all GCs and candidates. For comparison we also include the previous classifications from the RBC and C09. For consistency we have tried to keep our classifications similar to those used in the RBC. If we have no reason to reclassify the sources, we keep the original classifications (where available from C09, which were found to agree best with our classifications, otherwise from the RBC). The classifications used are: \\
\\
\textit{1: old globular cluster:} extended and has colours consistent with the Milky Way's GCs. Its velocity is confirmed from previous work (K07, C09 or the RBC) to be consistent with being in the M31 GC system, or the object is confirmed from high resolution \textit{HST} images. \\
\textit{2: candidate cluster:} not confirmed, but previously proposed as being a cluster or candidate and is found here to be extended (or have uncertain stellarity). Candidate is sub-divided, depending on whether its colours are consistent with being an old cluster (\textit{21}), consistent with being a young cluster (\textit{23}) or inconsistent with being a cluster (\textit{2}). \\
\textit{3: young cluster:} has colours consistent with being young. If previously classified, may appear as a resolved asterism in K, but looks like a cluster in the SDSS images. \\
\textit{4: background galaxy:} previously classified from spectroscopy by C09 or identified from our visual examination. \\
\textit{5: HII region:} from previous classification of C09. \\
\textit{6: stellar source:} object appears to be a single stellar source or a previously unresolved asterism of stellar sources. \\

The total number of sources of each class are shown in table \ref{tab:compare_class}. For reference we include whether these objects were previously classified as: clusters or candidates in the RBC (RBC 1 and RBC 2 respectively); clusters or candidates by K07 (K07 A and K07 B$/$C respectively); old or young clusters by C09 (C09 old and C09 young respectively). It can be seen that we have reclassified 10 previously confirmed clusters in the RBC as likely stellar sources. We also reclassify 6 of these objects as candidate clusters, as we are uncertain of they nature, or they lack spectroscopic confirmation. Some of the candidate clusters in the RBC are confirmed to be old or young clusters. This is based on the new spectroscopic confirmations by C09. We are also able to classify many of the candidate clusters in the RBC as stars. In most cases we find good agreement with the new classifications of C09. Their catalogue includes fewer objects because they do not provide classifications for the whole GC system. All objects classed as young clusters by C09 are retained in our classification. 

We have reclassified many of the confirmed clusters from K07 as likely stellar sources. We have also been able to classify nearly half of the cluster candidates from this catalogue as being stellar. We believe this is due to our improved spatial resolution compared with the images used for this previous catalogue. We identify and remove 8 objects from the catalogue of K07 which are within 2$\arcsec$ of another previously identified object in the RBC and we believe are now duplicated in the RBC. A further 5 objects from the catalogue of K07 appear to be associated with objects in the catalogue of C09. The names for these objects in table 1 are the combination of their identifications in each catalogue. 

\section{Properties of confirmed GCs}

\begin{figure}
 \includegraphics[height=85mm,angle=270]{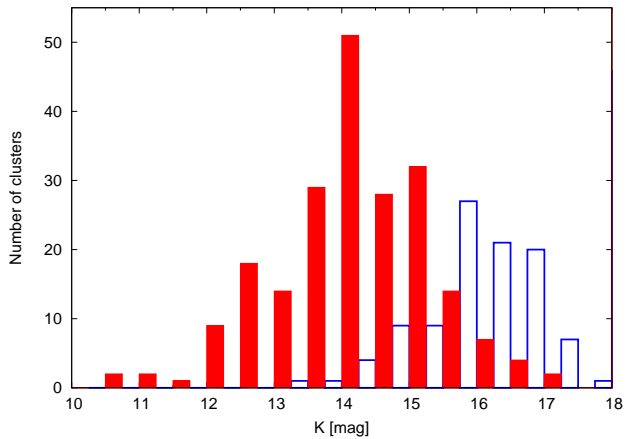}
 \caption{K-band GCLF for all sources classed as confirmed GCs (solid,~red) and young clusters (open,~blue).}
 \label{fig:gclf}
\end{figure}

Figure \ref{fig:gclf} shows the GC Luminosity Function (GCLF) for all confirmed GCs (solid bars) and young clusters (open bars) with K-band photometry. These clusters are not corrected for extinction. However, extinction is not very significant in the K-band where the maximum correction for the most extreme case of B037 is only 0.5~mag (the width of the bins used). The peak of the GCLF is found to be at K$\sim$14.2~mag. The K-band luminosity of a cluster is a useful estimate of its mass. This is because, in addition to being less effected by extinction, the K-band mass to light ratio ($M/L$) is less effected by metallicity than optical bands. The mass to light ratio of a 12~Gyr cluster in the K-band has previously been estimated to be 0.9$<$\textit{M/L}$<$1.3 for metallicities in the range 0$>$[Fe/H]$>$-2 \citep{Bruzual03,Forbes08}. To estimate the peak mass of the old GCs in M31, we assume a K-band $M/L$ ratio of 1.1 for all clusters (as the metallicities are not known for all of the clusters). At the distance of M31 \citep[780~kpc:][]{McConnachie05} and assuming the K-band magnitude of the sun to be M$_{\rm{Ks}\odot}$=3.29~mag [M$_{\rm{K}\odot}$=3.33: \citet{Cox00}; M$_{\rm{Ks}\odot}$=M$_{\rm{K}\odot}$-0.04: \citet{Carpenter01}], this implies a peak mass of M$_{peak}\sim3\times10^{5}\rm{M}_{\odot}$. This is slightly higher than that found for Milky Way GCs \citep[e.g.][]{Cohen07}. However, this difference is relatively small compared with the expected uncertainty in the peak mass. This is due to errors in accurately estimating the peak in the GCLF combined with errors on the distance to M31 and the value used for the mass to light ratio. 

For the fainter GCs, it is likely that masses estimated from their integrated K-band luminosities are less accurate due to stochastic effects. Stars at the tip of the red giant branch at the distance of M31 are expected to reach magnitudes of K=17.5 \citep{Ferraro00,Tabur09}. While stars this bright are relatively rare, the integrated light of some fraction of these faint clusters can therefore be dominated by a relatively low number of these stars. It is also likely that some of the faintest clusters in M31 are missing from our catalogue. These clusters should be detected in our data. However, identifying these faint clusters in front of M31 would be very difficult. 

As expected the proposed young clusters peak at fainter magnitudes than the old GCs. Some of these clusters are found to be relatively bright, reaching luminosities similar to the peak of the GCLF. This suggests they are more massive than typical young open clusters in the Milky Way. While this is in agreement with the conclusions of other work \citep[e.g.][]{Fusi_Pecci05}, it should be noted that our conclusions based on this K-band luminosity are limited. The $M/L$ ratio of these clusters is likely to be significantly lower than the $M/L$ ratio for the older clusters in the galaxy. Also stochastic effects in these young and faint clusters are likely to be significant in the K-band. 

\subsection{The structure of M31 GCs}

\begin{figure*}
 \includegraphics[height=155mm,angle=270]{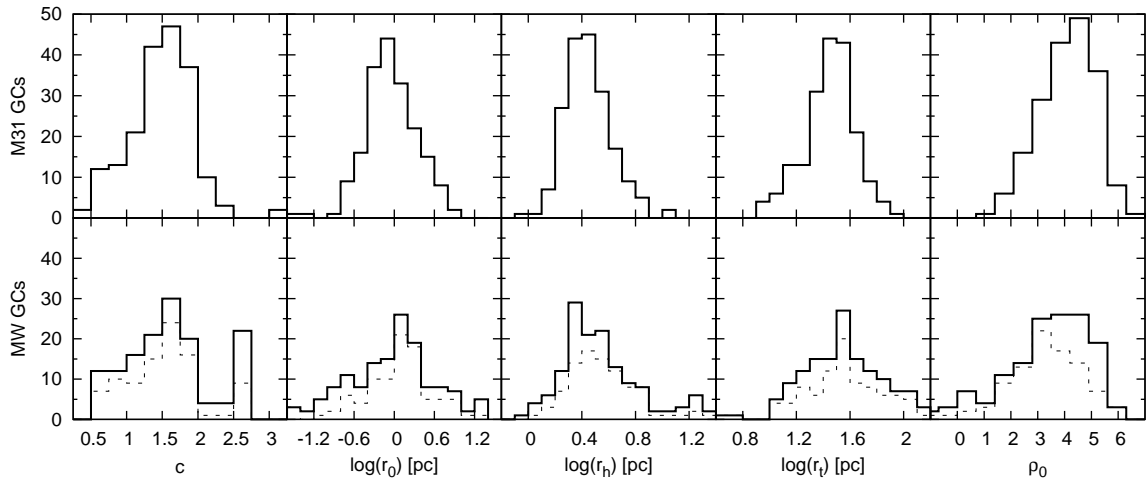}
 \caption{Structural parameters of M31 GCs (\textit{top}) and Milky Way GCs (\textit{bottom}). The dashed line indicates the GCs in the Milky Way over a similar galactocentric radius.}
 \label{fig:param}
\end{figure*}

The excellent spatial resolution of these WFCAM images allows us to investigate the structure of M31's old GCs. This can be done by fitting PSF convolved King models to their profiles. The results of this fitting to 239 \textit{class 1} clusters from the RBC are already presented in \citet{Peacock09} (hereafter P09). For convenience, we have included these structural parameters, where available, in table 1. Here we review these data and use it to consider the structure of M31's GC system. 

From our new classifications it is found that 186 of the 239 clusters studied by P09 are confirmed as old GCs. We have also determined the structural parameters for an additional 27, newly confirmed, GCs with WFCAM images. These parameters were determined using exactly the same method as P09. This results in a sample of the 213 old GCs with structural parameters (representing over half the confirmed GCs identified in M31). The reliability of these structural parameters, was investigated by P09 and they were found to be consistent with those obtained for 33 clusters whose parameters had already been measured using higher spatial resolution \textit{HST} images \citep{Barmby07}. They also found that the structural parameters obtained from fitting different WFCAM images of the same GC were consistent. However, the reliability of the parameters was found to decrease significantly for clusters with K$>$15~mag. We note that this decrease in reliability for the fainter clusters is in agreement with the signal to noise limit proposed by \citet{Carlson01}. The half light radii of the clusters was found to be the most reliable parameter \citep[as found by other studies e.g.][]{Kundu98}. Errors are not available for these parameters, but we believe them to be relatively reliable for most clusters with K$<$15~mag. 

Figure \ref{fig:param} shows the concentration [$c=$log$(r_{t}/r_{0})$], core radii ($r_{0}$), half light radii ($r_{h}$), tidal radii ($r_{t}$) and core density ($\rho_{0}$) of M31's old GCs. For comparison the same parameters for the Milky Way's GCs are shown on the bottom row \citep{Harris96}. Before comparing these populations there are two important differences which need to be considered. Firstly, the M31 parameters are based on the K, rather than the V-band luminosity of the clusters. As a result we expect offsets in parameters such as the core luminosity density due to the different mass to light ratios. However, this is unlikely to have a significant effect on the size of the clusters \citep{Cohen07}. Secondly, due to a lack of WFCAM data, the sample of M31 GCs does not include the most central or distant GCs in the galaxy. It can be seen from figure \ref{fig:param} that there are a lack of core collapsed GCs in our sample compared with the Milky Way. This can be partially explained by the exclusion of the innermost GCs (where most core collapsed GCs are located in the Milky Way). Taking this into account, we still find fewer of these clusters than expected. Potentially some faint core collapsed clusters may be missed by GC surveys as they would be the most difficult clusters to resolve. However, it is likely that these clusters are present, but have their concentrations underestimated. This is because their core radii will be much smaller than the PSF of our images, making it very hard to deconvolve and measure them. Comparison of the core radii of the clusters does show that we are missing, or overestimating, the core radii of some of the very smallest core radii clusters. A lack of core collapsed clusters can also be seen in similar profile fits to \textit{HST} images of Cen~A clusters \citep{Jordan07}. Allowing for these effects, we see no strong evidence for differences between the structure of the old GCs in M31 and Milky Way. 

\begin{figure}
 \includegraphics[height=85mm,angle=0]{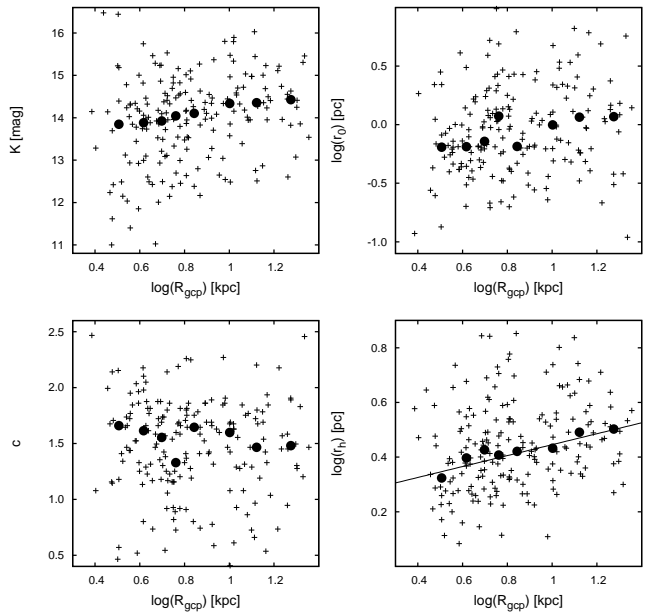}
 \caption{Properties of M31's GCs (crosses) as a function of projected distance to the center of M31 ($R_{gcp}$). Bold points show the median values for clusters binned on $R_{gcp}$. The line demonstrates the relationship found by \citet{Barmby07} between r$_{h}$ and $R_{gcp}$. } 
 \label{fig:Rgcp_param}
\end{figure}

In the Milky Way, it is known that some GC properties are related to position in the Galaxy. In figure \ref{fig:Rgcp_param} we plot the structural parameters of M31's GCs as a function of their projected galactocentric radius ($R_{gcp}:$ taken from the RBC). To help identify potential trends in these data, we have binned the clusters into groups of 25 and determined the median value of their parameters for each group (bold points). It can be seen that the luminosity of the GCs appears to decrease slightly with $R_{gcp}$. This is mainly driven by a deficit of faint clusters in this region combined with the most massive clusters being centrally located. The relatively low number of faint clusters in the central regions is likely due to selection effects, as it is very difficult to identify these clusters projected against the dense central regions of M31. It is therefore likely that some of these clusters are missing from our catalogue. Selection effects can not explain the lack of very massive clusters in the outer regions. However, there are very few of these very bright clusters. The distribution also suggests that more central clusters have smaller, more concentrated cores. This relationship may be expected from the evolution of the GC system as central GCs are expected to evolve more quickly due to greater interactions with their host galaxy. This effect is observed in the Milky Way's GCs \citep{Djorgovski94}. However, we again caution that this is a weak trend, and that the same selection effects could potentially prejudice us against extended, low density clusters in the inner regions. 

The half light radius of the clusters can be seen to increase with $R_{gcp}$. This has previously been observed for a smaller number of GCs in M31 but over a greater range of $R_{gcp}$ by \citet{Barmby07}. The line included in this plot is the relationship found by \citet{Barmby07} not a fit to our data. This demonstrates the excellent agreement between the trend they identify and that found here for a larger number of clusters. A similar trend is also found for GCs in the Milky Way \citep{vdBergh91,Djorgovski94} and in Virgo cluster galaxies \citep{Jordan05}. Unlike other cluster sizes, the half light radius of a cluster is thought to be largely unaffected by evolution. Therefore this relationship may be related to the properties of the globular cluster system at the time of formation.

\section{Conclusions}

Our final catalogue includes 416 old GCs. Where detected, we provide self consistent \textit{ugriz} and K-band photometry for the proposed clusters and candidate clusters. We note the difficulty in providing accurate photometry for some of these clusters due to the complex background of M31. We highlight the need for good spatial resolution in order to remove contamination from non cluster light when obtaining integrated magnitudes. Where available, we find our photometry to be consistent with that previously published. From our multicolour photometry, we confirm the population of very blue clusters identified previously. We show that these colours are consistent with their spectroscopic classification by C09 as young clusters. We note that many of these clusters look like resolved asterisms in our K-band images. However, some of these are confirmed by HST images to be genuine clusters. Higher spatial resolution optical images than available here are required in order to confirm their nature as genuine young clusters. 

We have identified that many of the confirmed clusters from \citet{Kim07} are likely stellar sources (we retain only 27 of their 111 confirmed clusters as old clusters). We also identify 10 confirmed clusters in the RBC as likely stellar sources. While we have considered the classifications from K07 and the RBC separately in this paper, we caution that \textit{all} of the objects confirmed by K07 to be clusters are included as confirmed clusters in the current version of the RBC (v3.5). We also provide new classifications for many of the cluster candidates proposed by this previous work. We identify many of these candidates to be stars and reduce the number of unclassified candidate clusters to 357. 

Taking extinction and selection effects into account, we find both the colours and structure of the old M31 GC system are consistent with the Milky Way's. We note a potential lack of both core collapsed and very extended GCs in our M31 sample. We caution that some (or all) of this effect may be due to selection effects in identifying these clusters, or difficulties in accurately measuring their parameters, rather than an intrinsic difference in the populations.

\section*{Acknowledgements}

We thank the referee of this paper, Michele Bellazzini, for his prompt, constructive, and helpful comments. We would like to thank Mark Rawlings, Andy Adamson and Mike Irwin from the UKIRT/CASU for obtaining these WFCAM observations and assisting with questions related to the WFCAM pipeline. We would also like to thank Sebastian Jester for his advise regarding the SDSS and Brian Warner and Alan Shafter for useful communication regarding novae.

\label{lastpage}

\end{document}